\documentclass[conferences]{IEEEtran}
\IEEEoverridecommandlockouts
\usepackage{cite}
\usepackage{amsmath,amssymb,amsfonts}
\usepackage{algorithmic}
\usepackage{graphicx}
\usepackage{textcomp}
\usepackage{xcolor}
\usepackage{bm}
\usepackage{lineno,float,amsfonts,amsmath,epsfig,graphics,booktabs,multirow,cite,stfloats}
\usepackage{amsthm,mathrsfs}
\usepackage{url}
\usepackage{colortbl}
\usepackage{subfigure}
\usepackage{comment}

\usepackage{fancyhdr}
\fancypagestyle{plain}{
  \fancyhf{} 
  \fancyhead[L]{\textcolor{blue}{This work has been submitted to the IEEE for possible publication. Copyright may be transferred without notice, after which this version may no longer be accessible.}} 
}

\usepackage[hidelinks]{hyperref}

\begin{document}

\title{Exploring the Interpretability of Forecasting Models for Energy Balancing Market}
\author{Oskar Våle, Shiliang Zhang,~\IEEEmembership{Member,~IEEE}, Sabita Maharjan,~\IEEEmembership{Senior Member,~IEEE}, Gro Klæboe
\thanks{\textcolor{blue}{This work has been accepted by \textit{Artificial Intelligence Science and Engineering}. DOI: 10.23919/AISE.2025.000020. This is a pre-print version.}\\Oskar Våle, Shiliang Zhang and Sabita Maharjan are with  Department of Informatics, University of Oslo, Oslo, Norway (e-mail: oskar.vaale@gmail.com, \{shilianz, sabita\}@ifi.uio.no). Gro Klæboe is with Department of Electric Energy, Norwegian University of Science and Technology, Trondheim, Norway (e-mail: gro.kleboe@ntnu.no). This work was supported via the grant ``Privacy preserving Transactive Energy Management (PriTEM)'' funded by UiO:Energy Convergence Environments.\\ \textcolor{blue}{© 2025 IEEE.  Personal use of this material is permitted.  Permission from IEEE must be obtained for all other uses, in any current or future media, including reprinting/republishing this material for advertising or promotional purposes, creating new collective works, for resale or redistribution to servers or lists, or reuse of any copyrighted component of this work in other works.}}
}

\maketitle

\begin{abstract}
The balancing market in the energy sector plays a critical role in physically and financially balancing the supply and demand. Modeling dynamics in the balancing market can provide valuable insights and prognosis to power grid stability and secure energy supply. While complex machine learning models can achieve high accuracy, their ``black-box'' nature severely limits the model interpretability. In this paper, we explore the trade-off between model accuracy and interpretability for the energy balancing market. Particularly, we take the example of forecasting manual frequency restoration reserve (mFRR) activation price in balancing market using real market data from different energy price zones. We explore the interpretability of mFRR forecasting using two models: (i) extreme gradient boosting machine (XGBoost) and (ii) explainable boosting machine (EBM). We also integrate the two models, and we benchmark all the models with a baseline naïve model. Our results show that EBM provides forecasting accuracy comparable to XGBoost while yielding a considerable level of interpretability. Our analysis also underscores the challenge of accurately predicting the mFRR price for the instances when the activation price deviates significantly from the spot price. Importantly, EBM's interpretability features reveal insights into non-linear mFRR price drivers and regional market dynamics. Our study demonstrates that EBM is a viable and valuable interpretable alternative to complex black-box AI models in the forecast for the balancing market.
\end{abstract}


\section{Introduction} \label{introduction}
A stable and reliable supply of electricity is crucial for the operation of any infrastructure in the modern society. The energy market plays a key role in maintaining grid stability by matching generation with consumption in real-time \cite{szabo2024impacts,CHENG2024114306}. Within this ecosystem, the balancing market is of critical importance for managing deviations, or power imbalances, that occur when planned operations differ from actual conditions \cite{ROSALESASENSIO2024113918,SIRIN2021112093} due to uncertainties in the demand and supply, \textit{e.g.}, renewable generation and vehicle charging and energy consumption~\cite{zhang2020privacy,zhang2021energy}. Vital balancing resources, \textit{e.g.}, Manual Frequency Restoration Reserve (mFRR), are activated to stabilize grid frequency when imbalance happens, and the cost to activate them is indicated by the activation prices \cite{10634929}. Accurate forecasts of the balancing market are highly valuable to market participants for managing risk of high energy costs and for Transmission System Operators (TSOs) to ensure grid stability and reliability \cite{UZUM2024109976,DBLP:journals/corr/abs-2503-03539}. Nevertheless, as the energy transition unfolds~\cite{duguma2023privacy}, modeling and predicting dynamics in the balancing market is increasingly challenging due the integration of more variable renewables in the power grid, whose generation is intermittent and with uncertainties by its nature.

The evolving energy system has spurred the use of sophisticated forecasting methods, with machine learning (ML) offering powerful tools for identifying complex patterns in energy data \cite{zhang2022extended,DBLP:journals/eninf/RajaperumalC25,wang2025ultra,zhang2025extended}. Advanced ML models, such as the widely-used gradient boosting machine XGBoost \cite{10902591}, have demonstrated state-of-the-art performance in many energy forecasting applications. However, their high accuracy often comes at the cost of interpretability and renders them ``black-box'' models \cite{10.1145/3725420}. In critical domains like the energy balancing market, understanding why a model makes a certain prediction is necessary for deriving insights, ensuring transparency, and designing risk-mitigation measures \cite{10858356}. This has fueled the growth of Explainable AI and the development of inherently interpretable ``glass-box'' models.

While literature exists on energy usage and price forecasting, studies specifically addressing the forecasting for the balancing market are rather sparse, more so when it comes to the accuracy-interpretability trade-off \cite{10738058}. Some recent work has highlighted the challenges in predicting balancing market prices and volumes, suggesting limitations in accurate predictions using publicly available data \cite{backe_predictions_2023}. In this paper, we address the gap by evaluating the trade-off between model prediction accuracy and interpretability for the balancing market. Particularly, we consider the forecasting of mFRR activation price, and we use the energy data of Norway's five price zones which are with dominant renewables in its energy generation. We build models for the forecast, and we compare the performance of the state-of-the-art XGBoost model with an inherently interpretable alternative, and the explainable boosting machine (EBM) \cite{jin2025exploring} which is designed to offer transparency through decomposable predictions without considerably compromising accuracy. To the best of our knowledge, we are the first to explore the accuracy-interpretability trade-off in forecasting balancing market dynamics.

We summarize the main contributions of this paper as: (i) evaluating the accuracy of the XGBoost model for mFRR activation price forecasting across five price zones (ii) evaluating the performance of an inherently interpretable EBM model and that of a stacked EBM-XGBoost ensemble, where we explore the model accuracy-interpretability trade-off, and (iii) leveraging EBM's interpretability to derive insights into non-linear price drivers and regional market dynamics. 

The remainder of this paper is organized as follows. In Section~\ref{section1}, we brief the background for mFRR activation price forecasting for the balancing market in Norway and the explainability of the forecast. In Section \ref{section2}, we describe about the models we considered in the forecast. In Section \ref{section:results}, we present the simulation setup and results, followed by analysis and discussions. Finally, Section \ref{section:conclusions} concludes the paper.

\section{Activation price prediction and explainability}\label{section1}

The energy balancing market aims to maintain real-time balance between electricity generation and consumption, ensuring operational security and system frequency stability \cite{ARUN2025115830}. In Norway, the balancing market is segmented into five distinct price zones from NO1 to NO5~\cite{zhang2025norwegian,zhang_2025_16794604}. When deviations occur, the Norwegian national TSO Statnett activates balancing energy reserves to restore balance \cite{hvidsten2025driving,nordic_balancing_model_bsp_2024}. In this paper, we look into the reserve of manual frequency restoration reserve (mFRR), and check the forecasting for it. The price paid for activating these reserves, in our example the mFRR activation prices, represents the marginal cost of this balancing action and is determined through the market \cite{MOUSAVI2025110576}. Such activation prices compensate Balancing Service Providers (BSPs) and form the basis for settling the financial deviations of Balance Responsible Parties (BRPs) \cite{11037417}. The mFRR activation price is inherently volatile and challenging to forecast accurately, as activations are driven by unforeseen deviations, real-time system needs, the magnitude of the imbalance, intermittence of production (especially by variable renewables), and transmission grid constraints \cite{nordic_balancing_model_memo_2024}.

As machine learning (ML) models become increasingly sophisticated and are deployed for such critical forecasting problems, the need for model interpretability becomes increasingly important \cite{molnar_interpretable_2022,shil2025cenergy3}. To this end, explainable AI (XAI) models are crucial not only for understanding the results, identifying biases and deriving insights, and but also for ensuring regulatory compliance - especially for critical domains like energy markets, and potentially also for discovering new domain-specific knowledge \cite{lundberg_unified_2017}. This study contrasts ``black-box'' models like XGBoost, which offer high performance but limited transparency, with ``glass-box'' models like the Explainable Boosting Machine (EBM), which provide inherent interpretability by design \cite{chen_xgboost_2016, nori_interpretml_2019}. For the structure, \textit{e.g.}, of the EBM model, it allows for direct inspection of how individual features contribute to a prediction, offering transparency not readily available in more opaque models.

\section{Models for mFRR activation price forecasting}\label{section2}

To investigate the trade-off between predictive performance and model interpretability for mFRR activation price forecasting, we employ a set of carefully selected machine learning models. The choice of models was driven by the need to benchmark against a simple approach, evaluate a high-performance ``black-box'' model, and thoroughly assess an inherently interpretable ``glass-box'' alternative. Below we describe these models along with their features, and the reason why we select them.

\subsection{Naïve model}

We adopt a naïve model where the forecast for the considered period is simply the last observed value. We use it as a baseline and we anticipate any sophisticated model should demonstrate significant improvement over this elementary approach.

\subsection{XGBoost}

XGBoost is a highly efficient and scalable implementation of the gradient boosting framework, which sequentially builds an ensemble of decision trees \cite{chen_xgboost_2016, hastie_elements_2017}. Each new tree is trained to correct the errors of the previous ensemble. XGBoost is widely recognized for achieving high forecasting accuracy on various tabular datasets and serves as a robust benchmark for complex, yet often opaque models. XGBoost combines the predictions from multiple and often weak models, and creates a strong overall model. In XGBoost, models are sequentially trained where each new model corrects the errors made by the previous ones, and the training relies on gradients to improve the predictions. The sequential training of models results in the power of gradient boosting in XGBoost.

The objective function for XGBoost consists of (i) the loss function to be minimized and (ii) the regularization term. The objective function can be expressed as
\begin{equation}
\mathcal{L} = \sum_{i=1}^{n} \ell(y_i, \hat{y}_i) + \sum_{k=1}^{K} \Omega(f_k)
\end{equation}
where $n$ is the number of instances, $y_i$ is the true label for instance $i$. $\hat{y}_i$ is the predicted value for instance $i$, $f_k$ represents the $k$-th tree in the ensemble, $\ell$ is the loss function measuring the difference between actual and predicted values, $K$ is the total number of trees. $\Omega(f_k)$ is the regularization term for the complexity of each tree, usually defined as

\begin{equation}
\Omega(f_k) = \gamma T + \frac{1}{2} \lambda ||w_k||^2 
\end{equation}
where $T$ is the number of leaves in tree $k$, $\gamma$ is a parameter that controls tree complexity, $\lambda$ is a regularization parameter, $w_k$ are the leaf weights. During the construction of trees, XGBoost optimizes the objective function using a greedy algorithm. The resultant prediction is combined as
\begin{equation}
\hat{y}i^{(t)} = \sum_{k=1}^{t} f_k(x_i) 
\end{equation}
where $f_k$ is the prediction from the $k$-th tree, $t$ is the current number of trees added to the model. To optimize the loss function, XGBoost uses first-order (gradient) and second-order (Hessian) derivatives of the loss function. Specifically, for a given prediction $\hat{y}_i$:

\begin{equation}
g_i = \frac{\partial \ell(y_i, \hat{y}_i)}{\partial \hat{y}_i} \quad \text{(gradient)}
\end{equation}
\begin{equation}
h_i = \frac{\partial^2 \ell(y_i, \hat{y}_i)}{\partial \hat{y}_i^2} \quad \text{(Hessian)}
\end{equation}

When finding the best split for a node, XGBoost computes the gain from splitting the data at a feature $j$
\begin{equation}
\text{Gain} = \frac{1}{2} \left( \frac{(\sum_{i \in \text{left}} g_i)^2}{\sum_{i \in \text{left}} h_i} + \frac{(\sum_{i \in \text{right}} g_i)^2}{\sum_{i \in \text{right}} h_i} - \frac{(\sum_{i=1}^{n} g_i)^2}{\sum_{i=1}^{n} h_i} \right) - \gamma
\end{equation}
and the final predictions from the model can be expressed as
\begin{equation}
\hat{y}i = \sum_{k=1}^{K} f_k(x_i) 
\end{equation}

While powerful, the resulting ensemble of many deep trees in XGBoost is often treated as a ``black-box,'' making it challenging to understand the precise reasoning behind its predictions.

\subsection{Explainable boosting machine (EBM)}

\begin{table*}[bp] 
\centering
\caption{Detailed metrics on validation dataset for mFRR activation price forecasts.}
\label{tab:metrics_validation}
\resizebox{0.8\textwidth}{!}{
\begin{tabular}{|l|rrr|rrr|rrr|rrr|}
\bottomrule
\multicolumn{13}{|c|}{Upwards mFRR activation price forecasts}\\

\toprule
 Model & \multicolumn{3}{c|}{Naïve} & \multicolumn{3}{c|}{XGBoost} & \multicolumn{3}{c|}{EBM} & \multicolumn{3}{c|}{Stacked} \\
Metric & MAE      & R2       & RMSE     & MAE      & R2       & RMSE     & MAE      & R2       & RMSE     & MAE      & R2       & RMSE     \\
\midrule
   no1  &  12.6252 &  -0.2389 &  45.4457 &   7.3639 &   0.5542 &  27.2610 &   7.2580 &   0.4453 &  34.9054 &   7.2976 &   0.4572 &  34.5303    \\
   no2  &  10.5150 &   0.5462 &  18.2591 &   4.6852 &   0.7921 &  12.4573 &   6.0851 &   0.7857 &  12.3654 &   4.7462 &   0.8081 &  11.7021    \\
   no3  &   8.4364 &   0.6105 &  17.0173 &   4.5368 &   0.8881 &   9.1221 &   8.2696 &   0.7323 &  13.5930 &   5.3416 &   0.8765 &   9.2351    \\
   no4  &   7.9726 &   0.6376 &  13.0649 &   5.5020 &   0.7159 &  11.5690 &   9.6987 &   0.5237 &  15.1106 &   6.3743 &   0.6862 &  12.2661    \\
   no5  &   9.2503 &   0.4964 &  20.8203 &   5.1325 &   0.7931 &  13.3444 &   6.4568 &   0.7969 &  12.2191 &   4.6942 &   0.8409 &  10.8138    \\
\bottomrule

\multicolumn{13}{|c|}{Downwards mFRR activation price forecasts}\\

\toprule
 Model & \multicolumn{3}{c|}{Naïve} & \multicolumn{3}{c|}{XGBoost} & \multicolumn{3}{c|}{EBM} & \multicolumn{3}{c|}{Stacked} \\
Metric & MAE      & R2       & RMSE     & MAE      & R2       & RMSE     & MAE      & R2       & RMSE     & MAE      & R2       & RMSE     \\
\midrule
   no1  &  10.0330 &   0.4807 &  22.1517 &   7.7712 &   0.5469 &  20.6908 &   9.5545 &   0.7208 &  14.1769 &   8.4438 &   0.7817 &  12.5360    \\
   no2  &  10.3121 &   0.4235 &  22.0653 &   7.6835 &   0.5785 &  18.8681 &   8.5140 &   0.7491 &  12.0150 &   7.2153 &   0.7808 &  11.2293    \\
   no3  &   8.0967 &   0.7082 &  12.9140 &   8.2782 &   0.7060 &  12.9621 &   9.3034 &   0.7440 &  12.2383 &   8.8894 &   0.7693 &  11.6170    \\
   no4  &   7.5834 &   0.6414 &  11.8028 &   8.1714 &   0.6799 &  11.1506 &   8.8006 &   0.7062 &  10.7175 &   8.7428 &   0.7081 &  10.6827    \\
   no5  &   8.4096 &   0.6562 &  13.7083 &   7.7181 &   0.6492 &  13.8469 &  10.4476 &   0.5913 &  14.9057 &   7.5929 &   0.6841 &  13.1047    \\
\bottomrule
\end{tabular}}
\end{table*}

EBM is designed to be inherently interpretable while still leveraging the power of gradient boosting as is in XGBoost, making them ideal for exploring the performance-interpretability trade-off. EBM is a generalized additive model (GAM) that incorporates gradient boosting to learn each feature's contribution \cite{nori_interpretml_2019}. The objective function in EBM follows a similar form to that of other boosting methods, but with a focus on producing additive models
\begin{equation}
\mathcal{L} = \sum_{i=1}^{n} \ell(y_i, \hat{y}_i) 
\end{equation}
where $n$ is the number of instances, $y_i$ is the true label for instance $i$, $\hat{y}_i$ is the predicted value for instance $i$, $\ell$ represents the loss function. The predicted value $\hat{y}$ from an EBM can be expressed as
\begin{equation}
\hat{y}_i = \sum_{j=1}^{p} f_j(x_{ij}) + \hat{y}_0 
\end{equation}
where $p$ is the number of features, $f_j$ is the contribution from the $j$-th feature, $x_{ij}$ is the value of the $j$-th feature for instance $i$, $\hat{y}_0$ is the global bias term or intercept. In EBM, each $f_j(x)$ can be represented as a smooth function, which allows for continuous relationships
\begin{equation}
f_j(x_{ij}) = f_j(\alpha_{j}) + f_j'(\beta_{j})(x_{ij} - \alpha_{j}) 
\end{equation}
where $\alpha_j$ is a reference point for feature $j$, $f_j$ is a parametric or non-parametric smooth function that maps the feature value to its contribution, $f_j'$ denotes the derivative or slope of this function at the point $\alpha_j$. EBMs maintain additivity across features, meaning that the impact of each feature on the prediction is independent of others. Therefore, the overall prediction can be easily interpreted as
\begin{equation}
\hat{y}_i = \hat{y}_0 + \sum_{j=1}^{p} f_j(x_{ij}) 
\end{equation}
The training process involves minimizing the loss function with respect to the model parameters using gradient boosting. For each boosting iteration $t$, the residual $r_i^{(t)}$ is calculated as
\begin{equation}
r_i^{(t)} = y_i - \hat{y}_i^{(t-1)}\label{eq:residual}
\end{equation}
and then fit the function $f_j$ to the residuals, and the prediction can be updated as
\begin{equation}
\hat{y}_i^{(t)} = \hat{y}_i^{(t-1)} + \eta f_j(x_{ij}) 
\end{equation}
where $\eta$ is the learning rate.

The primary advantage of EBM is its inherent interpretability. Global explanations are available through the ranked importance of each feature. Local explanations for individual predictions are derived by summing the contributions from each feature's shape function for the given instance.

\subsection{Stacked EBM+XGBoost model}

We implement a stacked EBM+XGBoost model to assess whether combining the EBM's predictions with an XGBoost model trained on its residuals can lead to further performance gains. In this ensemble model, we employ a two-level stacking architecture. The Level-0 base model is an EBM trained independently. The EBM's residual $r_i^{(t)}$ in (\ref{eq:residual}) on the training set is then used as the target variable for the Level-1 meta-learner. The meta-learner was an XGBoost model, trained using the same input features as the EBM but tasked with predicting these EBM residuals. The final prediction from the stacked model for a given instance is the sum of the EBM's output and the XGBoost meta-learner's predicted residual.

\section{Simulations and results}\label{section:results}
In this section, we describe the experimental setup, present the forecasting performance of the selected models in Section~\ref{section2}, and elaborate on the key findings, particularly focusing on the accuracy-interpretability trade-off and insights derived from the EBM.

\begin{table*}[htbp]
\centering
\caption{Detailed metrics on test dataset for mFRR activation price forecasts.}
\label{tab:metrics_test}
\resizebox{0.8\textwidth}{!}{
\begin{tabular}{|l|rrr|rrr|rrr|rrr|}
\bottomrule
\multicolumn{13}{|c|}{Upwards mFRR activation price forecasts}\\
\toprule
 Model &   &   Naïve &    &  & XGBoost &  &      &   EBM &       &  &   Stacked &      \\
Metric &     MAE &      R2 &    RMSE &     MAE &        R2 &      RMSE &     MAE &     R2 &    RMSE &     MAE &        R2 &      RMSE   \\
\midrule
   no1 &  14.973 &  0.3252 & 39.5797 &  8.9379 &    0.7051 &   26.1651 &  6.8112 & 0.7406 & 24.5423 &  6.7777 &    0.7404 &   24.5479   \\
   no2 & 16.0781 &   0.268 & 43.1884 &  4.8785 &    0.7703 &   24.1924 &  7.2108 & 0.7604 & 24.7102 &  4.7676 &    0.7703 &    24.195   \\
   no3 &  7.7871 &  0.2199 & 29.1914 &  4.9566 &    0.3656 &   26.3261 & 10.3383 & 0.2258 & 29.0811 &  4.8807 &     0.395 &   25.7086   \\
   no4 &  5.5549 &  0.0919 &   26.38 &  4.6359 &    0.1806 &   25.0596 &    6.85 & 0.1142 & 26.0545 &  4.7117 &    0.1607 &   25.3619   \\
   no5 & 12.0757 &  0.2787 & 35.6992 &  6.6637 &    0.6248 &   25.7489 &  7.7904 & 0.6444 & 25.0647 &   5.719 &    0.6455 &   25.0279   \\
\bottomrule

\multicolumn{13}{|c|}{Downwards mFRR activation price forecasts}\\

\toprule
 Model &   & Naïve   &    &  & XGBoost   &      &      &  EBM   &       &  & Stacked   &     \\
Metric &      MAE &      R2 &    RMSE &     MAE &        R2 &      RMSE  &     MAE &     R2 &    RMSE  &     MAE &        R2 &      RMSE  \\
\midrule
   no1 &  13.0629 &  0.4127 & 27.8561 &  6.9906 &    0.8297 &   14.9998  &   8.449 & 0.8338 & 14.8175  &  8.2953 &    0.8337 &   14.8245  \\
   no2 &  14.7361 &  0.2942 & 31.1818 &  8.8608 &    0.7722 &    17.715  & 10.4373 & 0.7372 & 19.0264  &   8.713 &    0.7018 &   20.2695  \\
   no3 &   6.0406 &  0.5103 & 13.2233 &  4.5606 &    0.7878 &    8.7041  &  6.2641 & 0.7279 &  9.8574  &   5.916 &     0.762 &    9.2182  \\
   no4 &    3.9839 &  0.4796 &  8.8056 &  4.1834 &    0.5976 &    7.7433 &  5.4147 &  0.548 &  8.2061  &  5.3485 &    0.5567 &    8.1269  \\
   no5 &   9.9953 &   0.444 & 21.7984 &  7.4878 &    0.8384 &   11.7531  &  7.7143 & 0.8349 &  11.878  &  5.2431 &    0.8621 &   10.8559  \\
\bottomrule
\end{tabular}}
\end{table*}

\begin{table*}[htbp]
\centering
\caption{Detailed metrics on test dataset for mFRR activation price forecasts after excluding predictions for times when $spot = target$ from the metric calculations. $N_{orig}$ and $N_{filter}$ indicate the number of original records and filtered records, respectively.}
\label{tab:nospot_metrics_test}
\resizebox{1\textwidth}{!}{
\begin{tabular}{|l|rrr|rrr|rrr|rrr|rrr|}
\bottomrule

\multicolumn{16}{|c|}{Upwards mFRR activation price forecasts}\\
\toprule
Area & N\textsubscript{orig} & N\textsubscript{filter} & removed & \multicolumn{3}{c|}{Naïve} & \multicolumn{3}{c|}{XGBoost} & \multicolumn{3}{c|}{EBM} & \multicolumn{3}{c|}{Stacked} \\
Metric & & & & MAE & R2 & RMSE & MAE & R2 & RMSE & MAE & R2 & RMSE & MAE & R2 & RMSE \\
\midrule
   no1 & 25504 & 7093 & 72.19\% & 18.3679 & 0.1230 & 56.0014 & 12.5790 & 0.4346 & 44.9635 & 11.3803 & 0.4362 & 44.9018 & 11.4199 & 0.4349 & 44.9531 \\
   no2 & 25504 & 5430 & 78.71\% & 21.1550 & 0.1047 & 60.6836 & 12.2021 & 0.3851 & 50.2912 & 12.8873 & 0.4045 & 49.4893 & 12.8415 & 0.3758 & 50.6672 \\
   no3 & 25504 & 5486 & 78.49\% & 12.6120 & 0.1027 & 54.9777 & 13.5348 & 0.1046 & 54.9177 & 13.8256 & 0.1462 & 53.6276 & 13.3269 & 0.1104 & 54.7415 \\
   no4 & 25504 & 6514 & 74.46\% & 11.1000 & 0.0929 & 47.3642 & 13.0213 & 0.0210 & 49.2051 & 14.1380 & -0.0105 & 49.9916 & 14.0279 & -0.0101 & 49.9805 \\
   no5 & 25504 & 8026 & 68.53\% & 18.7463 & 0.1546 & 56.0016 & 12.0879 & 0.4676 & 44.4421 & 12.2542 & 0.5042 & 42.8878 & 12.0161 & 0.4754 & 44.1149 \\
\bottomrule

\multicolumn{16}{|c|}{Downwards mFRR activation price forecasts}\\

\toprule
Area & N\textsubscript{orig} & N\textsubscript{filter} & removed & \multicolumn{3}{c|}{Naïve} & \multicolumn{3}{c|}{XGBoost} & \multicolumn{3}{c|}{EBM} & \multicolumn{3}{c|}{Stacked} \\
Metric & & & & MAE & R2 & RMSE & MAE & R2 & RMSE & MAE & R2 & RMSE & MAE & R2 & RMSE \\
\midrule
   no1 & 25504 & 9809 & 61.54\% & 11.8341 & 0.2610 & 22.8703 & 13.4706 & 0.2539 & 22.9789 & 11.7285 & 0.3601 & 21.2811 & 11.7665 & 0.3555 & 21.3570 \\
   no2 & 25504 & 9133 & 64.19\% & 14.3753 & -0.2346 & 28.4429 & 15.6579 & -0.1192 & 27.0815 & 17.7922 & -0.3128 & 29.3299 & 18.8288 & -0.6590 & 32.9711 \\
   no3 & 25504 & 9297 & 63.55\% & 5.9205 & 0.2990 & 12.0020 & 8.6507 & 0.1430 & 13.2710 & 8.7351 & 0.0556 & 13.9309 & 8.2264 & 0.2084 & 12.7542 \\
   no4 & 25504 & 11001 & 56.87\% & 4.7729 & 0.2509 & 10.1240 & 6.7444 & 0.1051 & 11.0655 & 6.1339 & 0.1926 & 10.5107 & 6.0740 & 0.2153 & 10.3622 \\
   no5 & 25504 & 8890 & 65.14\% & 9.8454 & 0.3452 & 17.7363 & 11.0979 & 0.4191 & 16.7051 & 10.9462 & 0.4083 & 16.8591 & 9.8746 & 0.3876 & 17.1524 \\
\bottomrule
\end{tabular}}
\end{table*}

\subsection{Dataset}
\label{subsec:data}
The data for this study is sourced from the Volue Insight API, covering the period from January 1, 2021, to March 15, 2025, from all the five Norwegian electricity price zones from NO1 to NO5. The dataset has a 15-minute temporal resolution. Key input features included the day-ahead spot price, forecasted consumption, forecasted hydropower and wind power production, forecasted residual load, forecasted heating and cooling needs, and cyclical temporal features (sine/cosine transformations of hour and month). The target variables to be forecasted by our models are the 8-hour ahead mFRR activation prices for both upwards and downwards regulation. Upwards and downwards regulation mean to increase and to reduce power generation, respectively, in restoring demand-supply balance. We use an expanding window cross-validation approach for model training and validation. To simulate a realistic forecasting scenario, when generating predictions for the test set, input features (\textit{e.g.}, consumption, production) are based on forecasts available at the time of prediction, rather than actual yet unseen observed values.

\subsection{Evaluation of model forecast accuracy}
\label{subsec:simulation_settings_performance}
In this section, we evaluate the forecast accuracy of all the models forementioned in Section~\ref{section2}. We use the metrics including Mean Absolute Error (MAE), Root Mean Squared Error (RMSE), and the Coefficient of Determination ($R^2$) in the evaluation. We select hyperparameters for XGBoost and EBM based on performance on a separate validation set by grid searching.

\begin{figure}[thb]
    \centering
    \includegraphics[width=0.65\columnwidth]{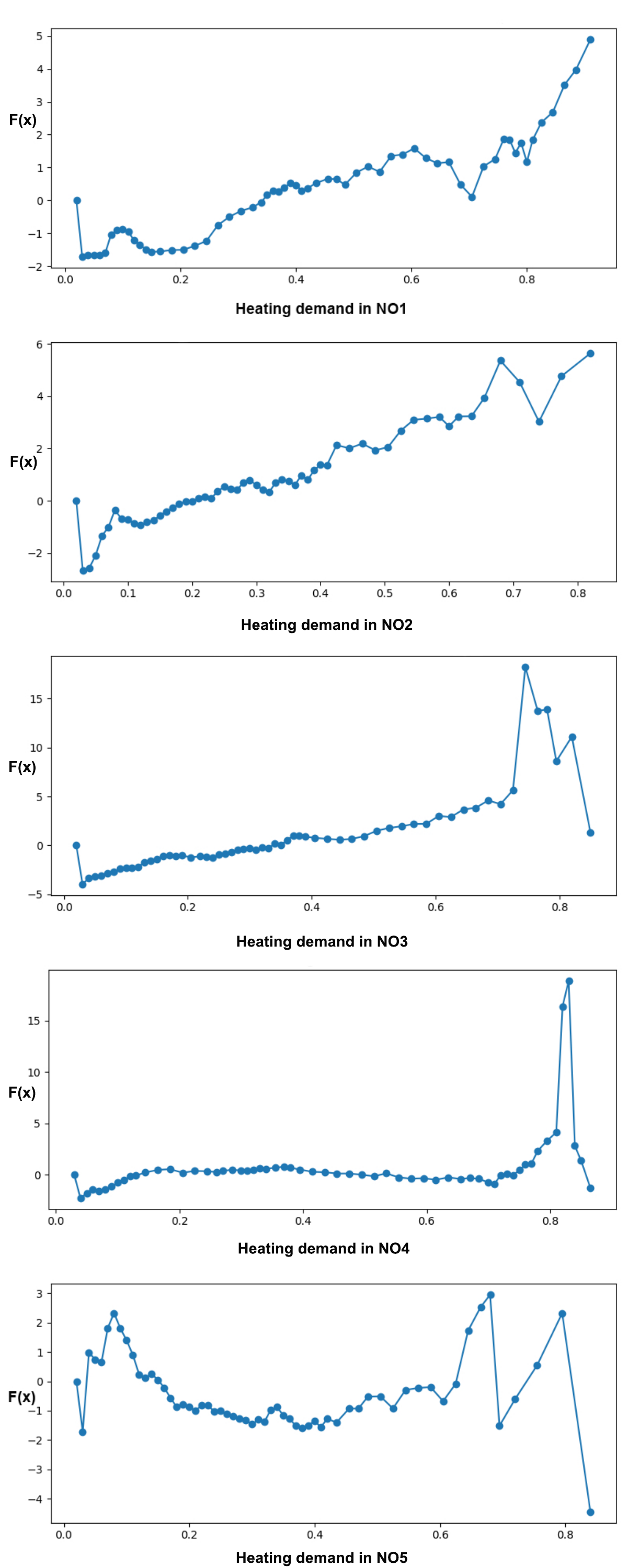}
    \caption{Learned EBM feature shape $f(x)$ for heating demand in upwards mFRR across price zones.}
    \label{fig:interpretability1}
\end{figure}

Table \ref{tab:metrics_validation} and \ref{tab:metrics_test} show the performance of the considered models on the validation and test dataset, respectively. We observe from the tables that both XGBoost and EBM significantly outperform the Naïve baseline, demonstrating their capability to capture the complex dynamics of mFRR prices. Notably, EBM achieves an accuracy comparable to the more complex XGBoost model. The stacked EBM+XGBoost model offers marginal improvements over the best standalone models, and this improvement differs across the price zones. This suggests that with the current feature set, the EBM or XGBoost alone captured most of the predictable signal for zones like NO5 for the downwards mFRR and NO4 for upwards mFRR, while the stacked model extracts more features in other zones and scenarios.

A significant novel finding is the challenge in forecasting periods where activation prices deviate from the day-ahead spot price, which is a dominant feature in determine the mFRR price. When models are evaluated exclusively on these ``deviation events'' (representing a smaller fraction of total instances, yet of critical importance in risk analysis for the balancing market), performance for all models including EBM and XGBoost drops substantially, as shown in Table \ref{tab:nospot_metrics_test}. For instance, $R^2$ values often approached zero or became negative, indicating that the models struggle to explain the variance in these specific yet critical price movements. This suggests that while those models can capture the general price level anchored by the spot price, predicting the timing and magnitude of significant departures remains a key difficulty with the current feature set.

\begin{figure}[htb]
    \centering
    \includegraphics[width=0.72\columnwidth]{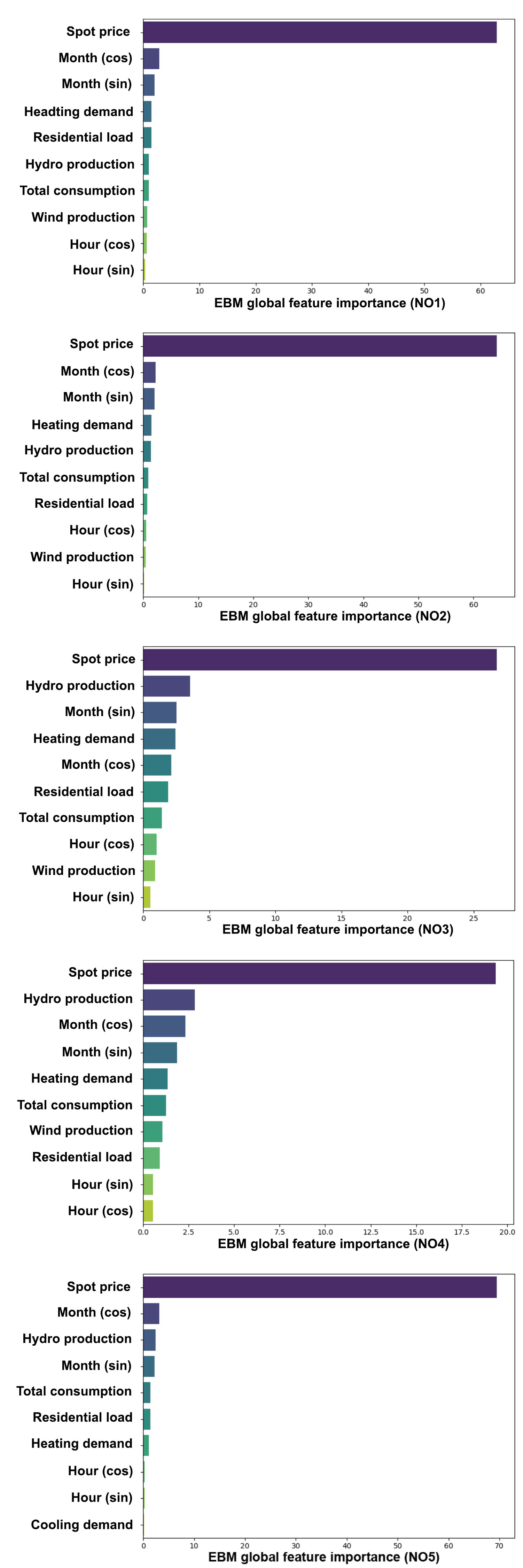}
    \caption{Global feature importance plots for each price zone for the upward mFRR price by the EBM model. ``Month (cos/sine)'' represents cosine/sine component of the cyclical encoding for the month of the year, and ``Hour (cos/sin) indicates cosine/sine component of the cyclical encoding for the hour of the day.}
    \label{fig:interpretability2}
\end{figure}

\begin{figure}[thb]
    \centering
    \includegraphics[width=0.65\columnwidth]{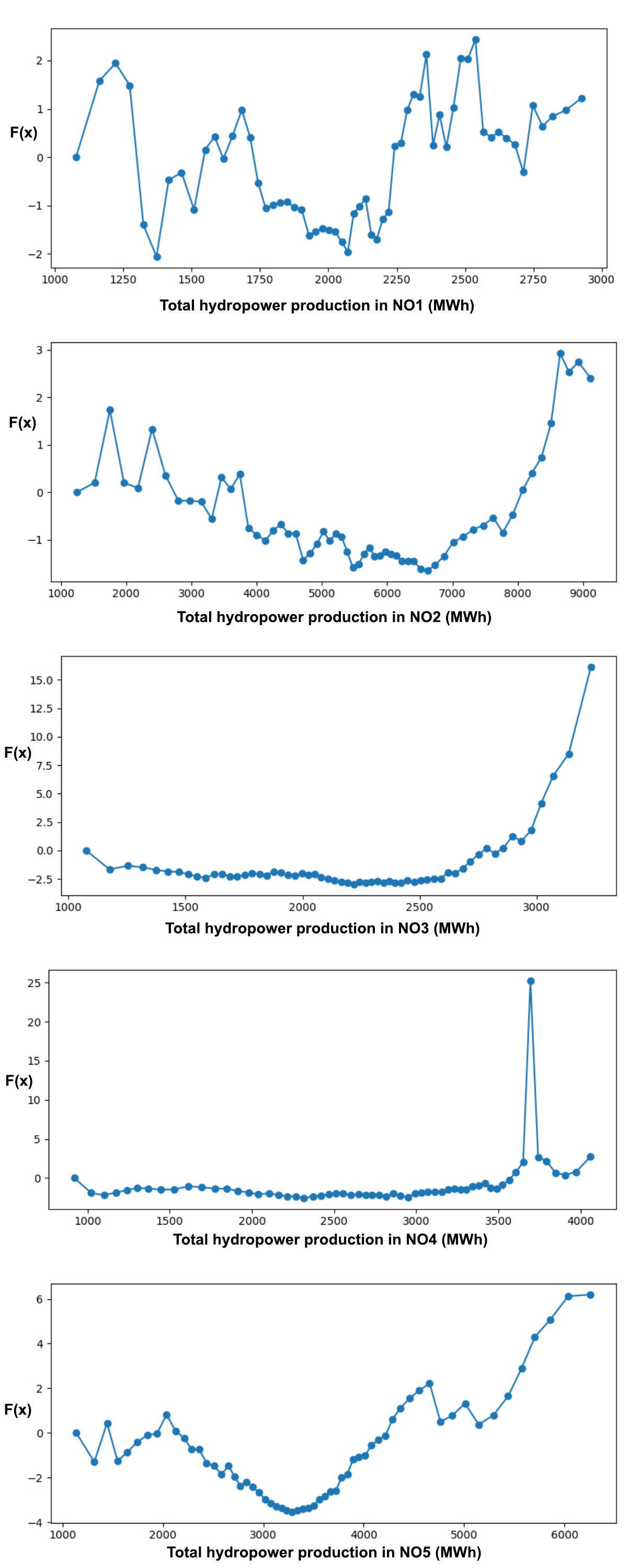}
    \caption{Learned EBM feature shape $f(x)$ for hydropower production in upwards mFRR across price zones.}
    \label{fig:interpretability3}
\end{figure}

\subsection{Result analysis and interpretability}
\label{subsec:result_analysis_discussions}

\begin{figure}[htb]
    \centering
    \includegraphics[width=0.75\columnwidth]{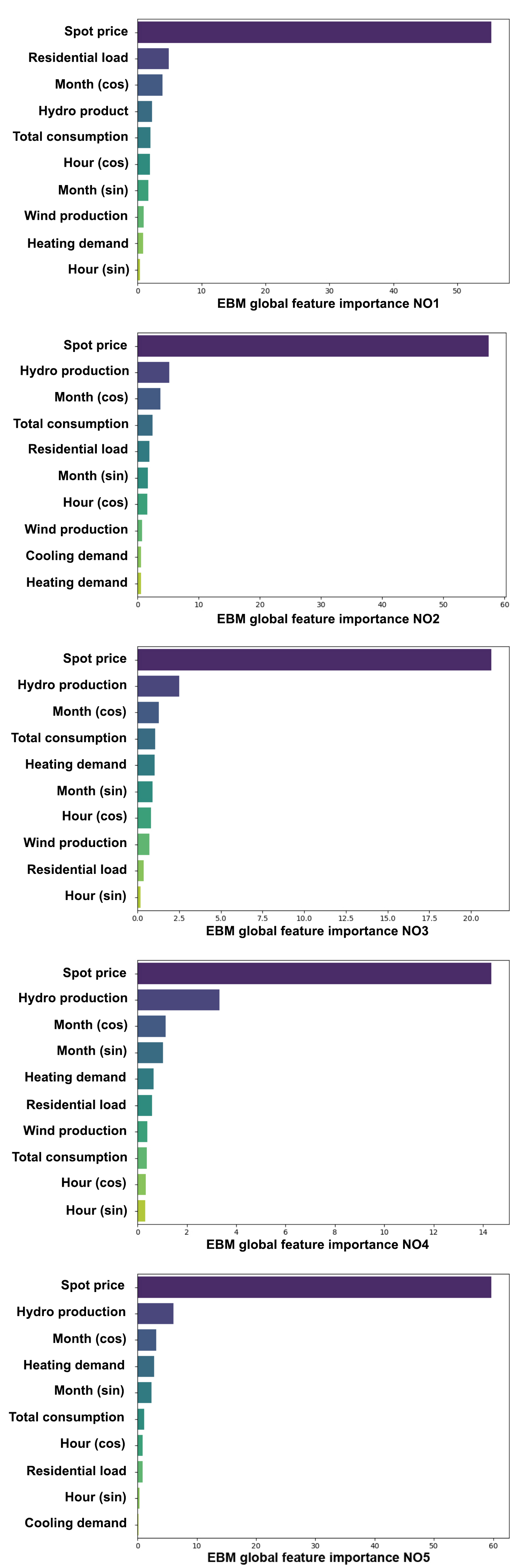}
    \caption{Global feature importance plots for each price zone for the downward mFRR price by the EBM model.}
    \label{fig:interpretability4}
\end{figure}

The EBM model's inherent interpretability provides valuable insights into the learned price drivers. In this section, we illustrate the interpretability of the gained EBM results shown from Fig.~\ref{fig:interpretability1} to \ref{fig:interpretability4}.  Global feature importance, measured by Mean Absolute Contribution (MAC), consistently identifies the day-ahead spot price as the most influential predictor across all zones, as indicated in Fig.~\ref{fig:interpretability2} and \ref{fig:interpretability4} for the upwards and downwards mFRR, respectively. There are also other important features that come behind the spot price, \textit{e.g.}, hydropower production (which consists of more than 80\% of the total power generation in Norway), heating demand since it is necessary in the winter time, and temporal features that reflect seasonal energy consumption and production patterns. We also observe that the latter features take effect to different levels for different energy price zones across Norway. \textit{E.g.}, hydropower production has a higher impact than seasonal factors in price zone NO3 and NO4, while not for NO1, NO2, and NO5. This corroborates the fact that NO3 and NO4 in Norway have sufficient production of hydropower, while are with less population than NO1, NO2, and NO5. This leads to the higher contribution of hydropower in NO3 and NO4 in local energy supply and its higher impact in the balancing market. While in NO1, NO2, and NO5, which are all in the southern part of Norway, they have a larger population that leads to higher heating demand in winter time, thus a higher impact of heating demand in the balancing market.

We show the relationship between upwards mFRR activation price and heating demand in Fig.~\ref{fig:interpretability1}, and that between upwards mFRR activation price and hydropower production in Fig.~\ref{fig:interpretability3}, respectively. Such a relationship is indicated by the shape function learned by the EBM model. Fig.~\ref{fig:interpretability1} displays near-linear positive relationship between mFRR and heating demand in price zone NO1 and NO2, reflecting that the large population and comparatively less energy generation in NO1 and NO2 induces high heating demand and uncertainties in the heating demand to be compensated by the balancing market.

Fig.~\ref{fig:interpretability3} shows that the impact of hydropower production does not have a significant impact on upwards mFRR activation price in NO3 and NO4, especially when the hydropower production is in a lower or mild level. This reflects that the sufficient hydropower generation in NO3 and NO4 does not add much uncertainties in the power grid that requests compensation by the balancing market. While in NO1, NO2, and NO5, the comparatively large population leads to more uncertainties in energy demand that needs to be compensated, and non-regular patterns can be observed for the relationship between mFRR activation price and hydropower production.

\section{Conclusions}\label{section:conclusions}

This paper addressed the trade-off between accuracy and interpretability for mFRR activation price forecast in the Norwegian energy market. Our results demonstrated that an inherently interpretable ``glass-box'' model, the explainable boosting machine (EBM), yields an accuracy of prediction comparable to the state-of-the-art ``black-box'' XGBoost model. 
The key implication of this work is that EBM can provide comparable accuracy to XGBoost while also yielding significant interpretability in balancing market predictions, providing direct insights into linear and non-linear price drivers and regional market dynamics that would otherwise be obscured. While this study underscores the potential of explainable AI, it also highlights that accurately predicting price deviations from the spot price remains challenging, largely due to the limitations of the current feature set. Future work should therefore focus on incorporating richer feature sets like real-time grid constraints, while also further exploring alternative model architectures.

\bibliographystyle{unsrt}
\bibliography{references}

\end{document}